\documentclass[reprint,aps,pra,showpacs,amsmath,amssymb,superscriptaddress]{revtex4-1}

\usepackage{color}
\usepackage{bm}
\usepackage{units}
\usepackage{graphicx}
\usepackage[caption=false]{subfig} 

\DeclareMathOperator{\tr}{tr}

\let\Re\relax
\let\Im\relax
\DeclareMathOperator{\Re}{Re}
\DeclareMathOperator{\Im}{Im}
\DeclareMathOperator{\Li}{Li}

\usepackage{braket}

\begin{document}

\title{Repulsive Casimir force between Weyl semimetals}

\author{Justin H. Wilson}
 \affiliation{Joint Quantum Institute and Condensed Matter Theory Center, Department of Physics, University of Maryland, College Park, Maryland 20742-4111, USA}
 \affiliation{School of Physics, Monash University, Melbourne, Victoria 3800, Australia}

\author{Andrew A. Allocca}
 \affiliation{Joint Quantum Institute and Condensed Matter Theory Center, Department of Physics, University of Maryland, College Park, Maryland 20742-4111, USA} 


\author{Victor Galitski}
 \affiliation{Joint Quantum Institute and Condensed Matter Theory Center, Department of Physics, University of Maryland, College Park, Maryland 20742-4111, USA}
 \affiliation{School of Physics, Monash University, Melbourne, Victoria 3800, Australia}

\begin{abstract}
Weyl semimetals are a class of topological materials that exhibit a bulk Hall effect due to time-reversal symmetry breaking.
We show that for the idealized semi-infinite case, the Casimir force between two identical Weyl semimetals is repulsive at short range and attractive at long range.
Considering plates of finite thickness, we can reduce the size of the long-range attraction even making it repulsive for all distances when thin enough.
In the thin-film limit, we study the appearance of an attractive Casimir force at shorter distances due to the longitudinal conductivity.
Magnetic field, thickness, and chemical potential provide tunable nobs for this effect, controlling the Casimir force: whether it is attractive or repulsive, the magnitude of the effect, and the positions and existence of a trap and antitrap.
\end{abstract}


\maketitle


In 1948, Casimir \cite{Casimir1948b} showed that  quantum fluctuations in the electromagnetic field cause a force between two perfectly conducting, electrically neutral objects.
This has since been extended to other materials \cite{Lifshitz:1956vv,Bordag:2001tt}.
Throughout this time, Casimir repulsion between two materials in vacuum has been a long sought after phenomenon \cite{Dzyaloshinskii1961,Boyer1974}.
There are principally four categories in which repulsion can be achieved: (i) modifying the dielectric of the intervening medium \cite{Dzyaloshinskii1961,Kenneth2002,Munday2009}, (ii) pairing a dielectric object and a permeable object \cite{Boyer1974} (such as with metamaterials \cite{Leonhardt2007,*Rosa2009}), (iii) using different geometries \cite{Boyer1968,Fulling2007,Levin2010}, and (iv) breaking time-reversal symmetry \cite{Tse2012,Grushin2011}.
In this paper, we are concerned with Casimir repulsion in identical time-reversal broken systems.
Specifically, we will study how Weyl semimetals with time-reversal symmetry breaking can exhibit Casimir repulsion.
The key ingredient to Casimir repulsion in this paper is the existence of a nonzero bulk Hall conductance $\sigma_{xy}\neq 0$, $\sigma_{xy}=-\sigma_{yx}$ \cite{Hosur2013}.

It is a general theorem that mirror symmetric objects without time-reversal symmetry breaking can only attract one another with the Casimir effect \cite{Kenneth2006}. 
This is understood with the Lifshitz formula \cite{Lifshitz:1956vv} where if we have two materials characterized by the two reflection matrices $R_1$ and $R_2$ and separated by a distance $a$ in a parallel plate geometry, we have 
\begin{align}
  E_c(a) =  \hbar\int \, \frac{d^2 k}{(2\pi)^2} \int \frac{d\omega}{2\pi}\, \tr \log [\mathbb I - R_1 R_2 e^{-2 q_z a}],\label{eq:Lifshitz}
\end{align}
where the trace is a matrix trace and $q_z = \sqrt{\omega^2 + k^2}$.
This integral generally yields an attractive force; however, if we break time reversal symmetry, obtaining antisymmetric off-diagonal terms in the reflection matrix $R_{xy}=-R_{yx}$ there is the possibility of Casimir repulsion \cite{Klimchitskaya2009}.
One candidate is a two-dimensional Hall material \cite{Tse2012}, and similarly, another is a topological insulator where the surface states have been gapped by a magnetic field \cite{Grushin2011,Rodriguez-Lopez2014}.
A Hall conductance does not guarantee repulsion; longitudinal conductance can overwhelm any repulsion from the Hall effect (although the magnetic field can lead to interesting transitions \cite{Allocca2014}), and a Hall effect that is too strong can suppress Kerr rotation and hence lead to attraction.
The latter case is an interesting phenomenon where ``more'' of a repulsive material can lead to attraction.

\begin{figure}[!tb]
\includegraphics[width=6cm]{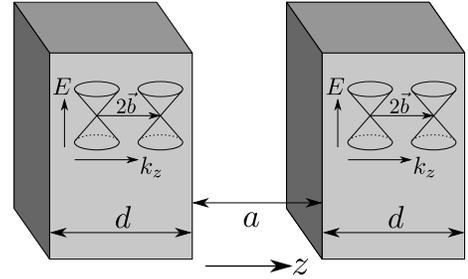}
\caption{The setup we will consider here is two Weyl semimetals separated by a distance $a$ in vacuum and with distance between Weyl cones $2\mathbf b$ in $k$ space (split in the $z$ direction). \label{fig:Picture-of-setup}}
\end{figure}

\begin{figure}
  \includegraphics[width=8.6cm]{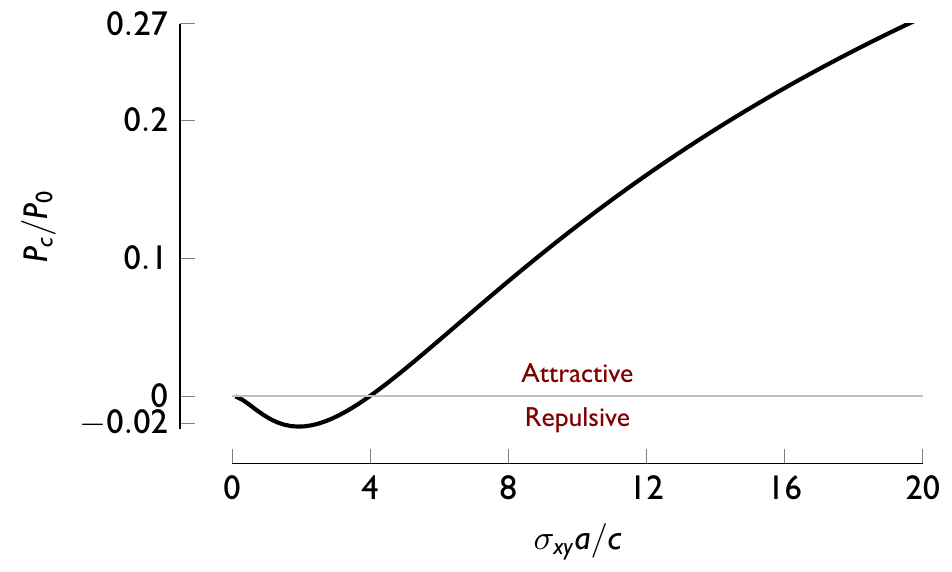}
  \caption{(Color online) The normalized Casimir force between two semi-infinite bulk Hall materials. Repulsion is seen for $\sigma_{xy} a/c \lesssim 4.00$. $P_0$ is the distant-dependent ideal Casimir force \cite{Casimir1948b}. For $\sigma_{xy}a/c\rightarrow\infty$, $P_c/P_0\rightarrow 1$.}
  \label{fig:SemiInfiniteBulkHall}
\end{figure}

\begin{figure}
  \includegraphics[width=8.6cm]{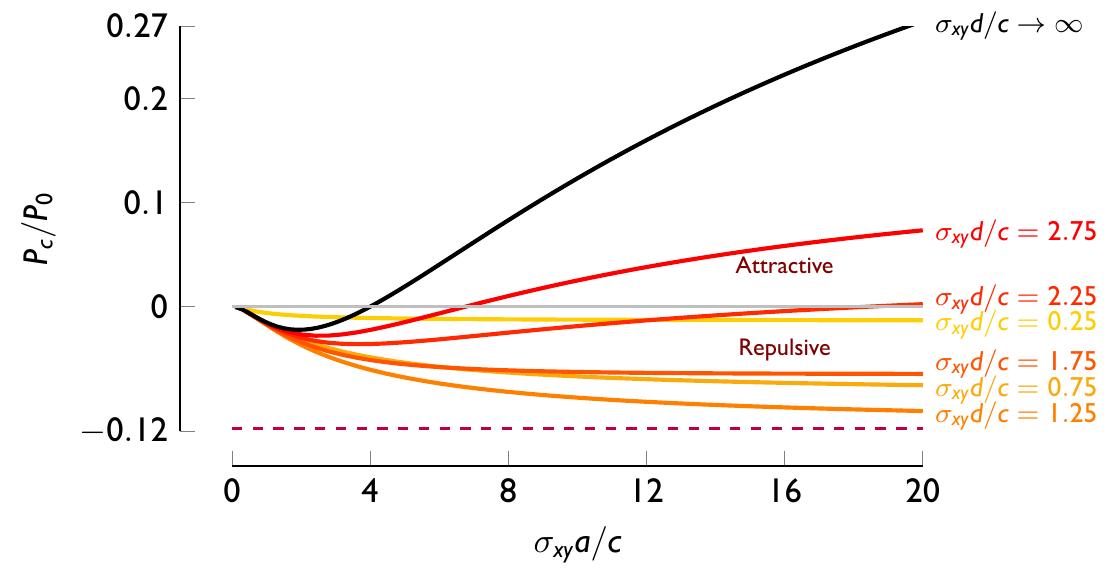}
  \caption{(Color online) A plot of the normalized Casimir force for various thicknesses of a bulk Hall material (idealized Weyl semimetal). It begins slightly repulsive for small $\sigma_{xy} d/c$, and as this increases, it becomes more repulsive until it reaches the maximum for a thin film material (the dashed line) at which point it increases to the semi-infinite limit.  $P_0 = - \frac{\hbar c\pi^2}{240 a^4}$, and $\sigma_{xy}=e^2 b/2\pi^2 \hbar$ is the bulk Hall response. 
  Figure~\ref{fig:thinfilmresults} takes into account more material properties. 
  \label{fig:ThickBulkHallmaterial}}
\end{figure}

\begin{figure*}[!tb]
\centering
\subfloat[]{
\includegraphics[width=5.7cm]{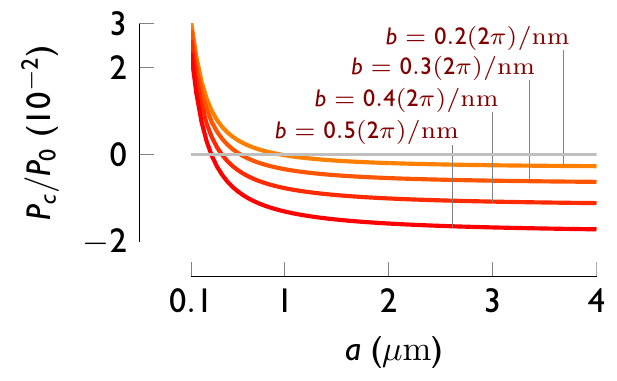}
\label{fig:WeylVaryb}}
\subfloat[]{
\includegraphics[width=5.7cm]{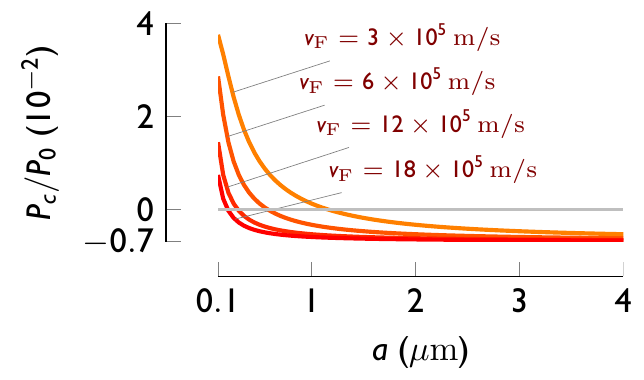}
\label{fig:WeylVaryvf}}
\subfloat[]{
\includegraphics[width=5.7cm]{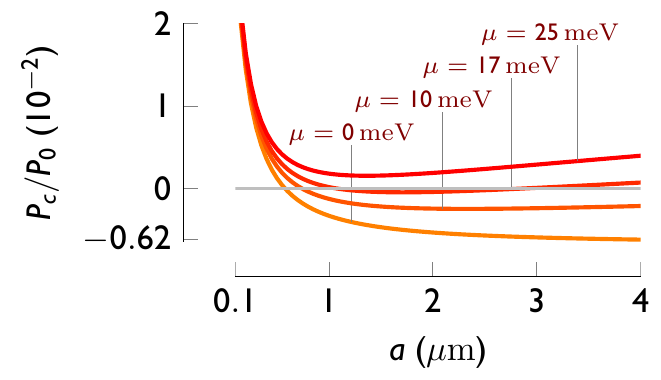}
\label{fig:WeylVarymu}}
\caption{(Color online) The Casimir force for a thin film Weyl semimetal taking into account low-energy virtual transitions in the band structure. An antitrap develops when the longitudinal conductivity overwhelms the Hall conductivity. In \protect\subref{fig:WeylVaryb} we compare different values of $b$ (or equivalently, changing the Hall effect), in \protect\subref{fig:WeylVaryvf} we compare different $v_{\mathrm F}$'s (the larger $v_{\mathrm F}$, the smaller $\sigma_{xx}$ is), and in \protect\subref{fig:WeylVarymu} we turn on a finite chemical potential which causes attraction at very long distances (and hence a trap). Even small chemical potentials have this property, but the trap is quite far out. Unless the parameter is varying, $a_0=\unit[1]{nm}$, $d=\unit[20]{nm}$, $b=0.3 (2\pi/a_0)$, $\Lambda=2\pi/a_0$, $v_{\mathrm{F}} = \unit[6\times 10^5]{m/s}$, and $\mu=0$.\label{fig:thinfilmresults}}
\end{figure*}

The material we are interested in is marginal in both the case of longitudinal conductance and in an overwhelming Hall effect: Weyl semimetals \cite{Hosur2013} with the Casimir setup seen in Fig.~\ref{fig:Picture-of-setup} and the resulting normalized Casimir pressure seen in Figs.~\ref{fig:SemiInfiniteBulkHall}--\ref{fig:thinfilmresults}.
These materials have linearly dispersive band structures characterized by Weyl nodes  with different chiralities and characterized by a chiral anomaly \cite{Aji2012}.
Clean Weyl semimetals at zero temperature have a zero dc longitudinal conductivity and optical conductivity $\Re[\sigma_{xx}]\propto \omega$ \cite{Hosur2012}.
Additionally, they exhibit a bulk Hall effect exemplified in the dc limit by an axionic field theory \cite{Zyuzin2012} where in addition to the Maxwell action, we have
\begin{align}
\label{eq:axionic-action}
  S_{A} = \frac{e^2}{32 \pi^2\hbar c} \int d^3 r\, dt \, \theta(\mathbf r, t)  \epsilon^{\mu \nu \alpha \beta} F_{\mu \nu} F_{\alpha \beta},
\end{align}
where $\theta(\mathbf r, t) = 2\mathbf b \cdot \mathbf r - 2 b_0 t$ and $2 \mathbf b$ is the distance between Weyl nodes in $\mathbf k$ space whereas $2 b_0$ is their energy offset, $e$ is the charge of an electron, $\hbar$ is Planck's constant, $c$ is the speed of light, $F_{\mu \nu}$ is the electromagnetic field strength tensor, and $\epsilon^{\mu \nu \alpha \beta}$ is the fully antisymmetric four-tensor.
Inversion symmetry breaking Weyl semimetals, on the other hand, do not exhibit a dc Hall effect \cite{Halasz2012} and therefore will not see the effects described in this paper.
The electrodynamics of this were investigated in Ref.~\cite{Grushin2012} where the authors even comment on the possibility for a repulsive Casimir effect.

The marginal nature of Weyl semimetals makes them prime candidates for tuning the Casimir force between attractive and repulsive regimes.
In constructed Weyl semimetals made of heterostructures of normal and topological insulators \cite{Burkov2011} an external magnetic field can control the Hall effect \cite{Chen2013a} and hence the repulsive effects.
Additionally, some of the first materials that have been predicted were pyrochlore iridates \cite{Wan2011,*Chen2012,*Witczak-Krempa2012}; these could also see a repulsion tunable with carrier doping or an additional magnetic field.

In a real material and experiment at finite temperatures, disorder and interactions should be taken into account, and in Weyl semimetals they lead to a finite dc conductivity \cite{Hosur2012,Burkov2011a,Burkov2011}.
We simulate this effect in the latter part of this paper by raising the chemical potential in our clean system, leading to intraband transitions that contribute to the longitudinal conductivity (in the dc limit these are singular contributions).

To begin, we consider two semi-infinite slabs of a Weyl semimetal ($z<0$ and $z>a$ to be precise), neglecting all frequency dependence to the conductivities by assuming the electromagnetic response is captured by Eq.~\eqref{eq:axionic-action}.
The result is just a material that is solely a bulk Hall material with current responses given by the Hall conductivity $\sigma_{xy} = e^2 b/2\pi^2 \hbar$.
This response can be encoded in the dielectric function so that $\epsilon_{xx}=\epsilon_{yy}=\epsilon_{zz}=1$ and $\epsilon_{xy}=-\epsilon_{yx}=i\sigma_{xy}/\omega$.
With this set up, if an incident wave $\mathbf k$ hits such a material it will break up into two different polarizations in the material $\mathbf k^{\pm}$ that satisfies $k_x^\pm=k_x$, $k_y^\pm=k_y$, and $(k_z^\pm)^2 = k_z(k_z\pm \sigma_{xy}/c)$. Additionally, the two elliptical polarizations $\mathbf D_\pm = \epsilon(\omega) \mathbf E_\pm$ are 
$\mathbf D_\pm \propto \tfrac{\omega}{c k^{\pm}} (k_z \pm \sigma_{xy}/c) \mathbf{e}_1 \mp i k_z^\pm \mathbf{e}_2$
where $\mathbf{e}_2$ is perpendicular to the plane of incidence and $\mathbf{e}_1= \mathbf{e}_2 \times \mathbf{k}^\pm$. 
Notice that for $k_z<\sigma_{xy}/c$, one of the polarizations is evanescent.

The incident and reflected polarizations can be broken up into transverse electric (TE) and transverse magnetic (TM) modes, and the reflection matrix $R(\omega,\mathbf k)$ just connects incident to reflected $(E^{\mathrm{TM}}_r, E^{\mathrm{TE}}_r)^T = R(\omega, \mathbf k)(E^{\mathrm{TM}}_0, E^{\mathrm{TE}}_0)^T$.
As shown in the Lifshitz formula Eq.~\eqref{eq:Lifshitz}, we need the imaginary frequency reflection matrix.
If we let $\omega\rightarrow i \omega$ and define $q_z^2 = \omega^2/c^2+k_x^2+k_y^2$, the reflection matrix for a semi-infinite slab of this bulk Hall material is
\begin{align}
 R_\infty(i q_z) = \frac1{ \sigma_{xy}/c}\begin{pmatrix}
    Q_- - \sigma_{xy}/c & - Q_+ +2 q_z \\
     Q_+ -2 q_z & Q_- - \sigma_{xy}/c 
 \end{pmatrix},
\end{align}
where we have defined for brevity $Q_{\pm} = \sqrt{2q_z(\sqrt{q_z^2+\sigma_{xy}^2/c^2} \pm q_z)}$ (the real frequency version of $R_\infty$ is found in the Supplemental Material \cite{supplement}).
Inspecting $R_\infty(i q_z)$, we see that the reflection matrix only depends on the ratio $c q_z/\sigma_{xy}$.
This dependence has implications for the Casimir force. 
After changing variables to solely $q_z$, we can inspect the Casimir pressure $P_c(a) = - \partial E_c/\partial a$, and we have an expression $P_c = \frac{2 \hbar c}{(2\pi)^2} \int d q_z\, q_z^3 \; g\!\left[\tfrac{q_z}{\sigma_{xy}/c}, 2 q_z a \right], \label{eq:semi-inf-force}$ with a function $g(\tfrac{q_z}{\sigma_{xy}/c}, 2 q_z a)$ written out in the Supplemental Material \cite{supplement} for completeness.
If we then change variables to $x = 2 a q_z$ and normalize by Casimir's original result for perfect conductors $P_0 = - \frac{\hbar c\pi^2}{240 a^4}$ \cite{Casimir1948b}, we can write the equation for the pressure as $P_c/P_0 = f(\sigma_{xy} a/c)$.

With this formulation, we plot normalized force $P_c/P_0$ as a function of $\sigma_{xy} a /c$ obtaining the single function seen in Fig.~\ref{fig:SemiInfiniteBulkHall}.
We see that for $\sigma_{xy} a/c \lesssim 4.00$ we obtain repulsion whereas for $\sigma_{xy} a/c \gtrsim 4.00$ we obtain attraction.
Thus, these similar materials trap each other at a fixed distance simply dependent on the Hall conductivity, $a_{\mathrm{Trap}} \approx \frac{4.00}{\sigma_{xy}/c}$.
If we insert the value of $\sigma_{xy}=e^2 b/2\pi^2\hbar$ into this expression, we find $a_{\mathrm{Trap}} \approx 860/b$. 
This means that if $1/b\sim O(\unit{nm})$, then $a_{\mathrm{Trap}}\sim O(\unit{\mu m})$ quite reasonable.

As the distance between the materials gets long, $P_c/P_0 \rightarrow 1$.
This behavior is markedly different from the thin film Hall case obtained by Tse and MacDonald in Ref.~\cite{Tse2012} and Rodriguez-Lopez and Grushin Ref.~\cite{Rodriguez-Lopez2014}.
They found a small [two-dimensional (2D)] quantum Hall effect implies a quantized and repulsive Casimir force at long distances.
In our case, we get attraction at long distances for a bulk Hall material independent of the magnitude of the Hall effect.
To resolve this seeming inconsistency, imagine a finite thickness of the bulk Hall material of thickness $d$, then the two-dimensional conductivity $\sigma_{xy} d$ diverges as $d\rightarrow\infty$, and in the case of a 2D quantum Hall plate with infinite Hall conductivity, the Casimir effect is attractive and approaches $P_0$.

To make this argument more precise, one can actually find the reflection matrix for a bulk Hall system of thickness $d$ and the result is (derivation of $R_d$ depends only on the axionic action Eq.~\eqref{eq:axionic-action} and can be found in the Supplemental Material \cite{supplement}, calculated for real frequencies)
\begin{align}
R_d(i q_z) & = \begin{pmatrix}
  R_{xx} & R_{xy} \\ - R_{xy} & R_{xx}
\end{pmatrix},
\end{align}
with
\begin{widetext}
\begin{align}
 R_{xx} & = -\tfrac12\tfrac{\sigma_{xy}}{c}( Q_- \sinh Q_+ d +\tfrac{\sigma_{xy}}{c}\cosh Q_+ d  - Q_+ \sin Q_- d -\tfrac{\sigma_{xy}}{c}\cos Q_- d)/D, \\
 R_{xy} & = -\tfrac12 \tfrac{\sigma_{xy}}{c}( Q_+ \sinh Q_+ d + 2 q_z \cosh Q_+ d  - Q_- \sin Q_- d - 2q_z \cos Q_- d)/D,
\end{align}
where
\begin{align}
 D  = (Q_+^2 +\tfrac12\tfrac{\sigma_{xy}^2}{c^2}) \cosh Q_+ d + (2 q_z Q_+ + \tfrac{\sigma_{xy}}{c} Q_-) \sinh Q_+ d + (Q_-^2 - \tfrac12 \tfrac{\sigma_{xy}^2}{c^2}) \cos Q_- d + (2q_z Q_- - \tfrac{\sigma_{xy}}{c} Q_+)\sin Q_- d.
\end{align}
\end{widetext}
It can be shown that $\lim_{d\rightarrow \infty} R_d(iq_z) = R_\infty(i q_z)$.

Similarly, in the limit of $d \rightarrow 0$, if we keep $\sigma_{xy}^{2\mathrm D} = \sigma_{xy} d$ constant, we obtain what was found in Ref.~\cite{Tse2012},
\begin{align*}
  \lim_{d\rightarrow 0} R_d(i q_z) = \frac1{1 + (\sigma_{xy}^{2\mathrm D}/2c)^2 } \begin{pmatrix}
    -(\sigma_{xy}^{2\mathrm D}/2c)^2 & -\sigma_{xy}^{2\mathrm D}/2c \\
    \sigma_{xy}^{2\mathrm D}/2c & -(\sigma_{xy}^{2\mathrm D}/2c)^2
  \end{pmatrix}.
\end{align*}
For the rest of our discussion, define $R_0(i q_z) = \lim_{d\rightarrow 0} R_d(i q_z)$ with $\sigma_{xy}^{2\mathrm D} = \sigma_{xy} d$ held constant.

With the correct limits identified, we first notice that we can write $R_d$ as a function of only two variables $R_d = R_d(c q_z/\sigma_{xy}, \sigma_{xy}^{2\mathrm D}/c)$.
Thus, by similar arguments to what we had for the semi-infinite case, the Casimir pressure $P_c = P_0 f(\sigma_{xy} a/c, \sigma_{xy}^{2\mathrm D}/c)$.

The limiting cases can be understood now by considering first Eq.~\eqref{eq:Lifshitz}.
The exponential constrains $q_z \sim 1/a$ and since technically the thin-film limit is $\lim_{q_z d\rightarrow 0} R_d(i q_z) = R_0(i q_z)$, we have that $d/a \rightarrow 0$.
In other words, the thin film limit is applicable when we are considering $d \ll a$.
The opposite limit is just when $q_z d \rightarrow \infty$, and by similar arguments, that means $d \gg a$ is when the semi-infinite case applies.
Both limits leave $\sigma_{xy} a/c$ and $\sigma_{xy} d/c$ unaffected (although in the thin film case $\sigma_{xy} a$ drops out whereas in the semi-infinite case $\sigma_{xy} d\rightarrow \infty$ has the same limit as $q_z d \rightarrow \infty$).

The thin film limit can be evaluated exactly \cite{Tse2012} and has the form
  $P_c^{\mathrm{TF}} = P_0 \frac{90}{\pi^4} \Re \{ \Li_4\left[(\sigma_{xy}^{2\mathrm D}/c)^2/(\sigma_{xy}^{2\mathrm D}/c+2 i)^2 \right]\}$
where $\Li_4$ is the polylogarithm of degree 4.
Note that this function has a minimum value of $P_c^{\mathrm{TF}} \approx -0.117 P_0$ representing how repulsive we can get.
For large enough $\sigma_{xy}^{2\mathrm D}/c$, the force does become attractive---corresponding roughly to when $(\sigma_{xy}^{2\mathrm D}/c)^2>\sigma_{xy}^{2\mathrm D}/c$ (i.e.\ when Kerr rotation is suppressed).

The cross-over between these regimes can be seen in Fig.~\ref{fig:ThickBulkHallmaterial}. 
As $\sigma_{xy}d/c$ is increased, the Casimir energy approaches the semi-infinite case.
However, for any finite $d$, each curve asymptotically approaches its thin-film value (and never goes lower than the minimum value represented by the dashed horizontal line in Fig.~\ref{fig:ThickBulkHallmaterial}).
This not only clearly connects our case to the previously known thin-film result, but also provides a theoretical justification for considering a thin-film limit $d\ll a$ with a two-dimensional conductance $\sigma_{\mu\nu} d$.

Until now the plates have been idealized. 
Using the thin-film limit illustrated above as a reference allows us to easily consider some of the effects of taking into account the full frequency response of the material.
Thus, we pick a $\sigma_{xy} d$ that is reasonably in the repulsive regime (for all distances) in order to analyze the effects of including some of the lowest-order frequency dependence into the conductivities.
We will mainly be interested in the effects of virtual vacuum transitions that are low in energy, which correspond to plates that are far apart from one another.
Thus, we will use the low-energy chiral Hamiltonian for a pair of Weyl nodes,
\begin{align}
  H_W = \pm \hbar v_{\mathrm F} \bm \sigma \cdot (\mathbf k \pm \mathbf b),
\end{align}
where $v_{\mathrm F}$ is the Fermi velocity and $\mathbf b$ is the position of the of Weyl node in $\mathbf k$ space.
The exact band structure will be important as the plates get closer although weighting will still be larger on the lower-energy modes.

To the conductivities, we fix $k_z$ and calculate two-dimensional conductivities using the Kubo-Greenwood formulation (see the Supplemental Material \cite{supplement} for details) then integrate the resulting expression over $k_z$ with a symmetric cutoff $\sigma_{\mu \nu}(i \omega) = \int_{-\Lambda}^\Lambda \frac{d k_z}{2 \pi} \tilde\sigma_{\mu \nu}(i \omega;k_z)$ \cite{Goswami2013}
where $\tilde\sigma_{\mu\nu}(i\omega;k_z)$ is the two-dimensional conductivity with $k_z$ fixed.
We evaluate this at an imaginary frequency to aid the Casimir calculations.

We perform this procedure at finite chemical potential $\mu$ and throw out terms that go to zero when the cutoff $\Lambda\rightarrow \infty$. 
This yields the conductivities,
\begin{multline}
 \sigma_{xx}(i\omega) = \frac{e^2}{12 \pi^2 \hbar v_{\mathrm F}} \left[ \tfrac 53 \omega + 2 \omega \log \left( \tfrac{v_{\mathrm F} \Lambda}{\omega} \right) \right. \\ \left.+ 4 \tfrac{\mu^2}{\hbar^2\omega} - \omega \log \left(1 + \tfrac{4 \mu^2}{\hbar^2\omega^2} \right)\right],
\end{multline}
and $\sigma_{xy}(i \omega) = \frac{e^2 b}{2 \pi^2 \hbar}$ is unchanged at this order.
Due to the linear dispersion of the Weyl nodes, we have a logarithmic cutoff dependence. 
Note that rotating to real frequencies we get the correct result for two Weyl nodes for $\Re[\sigma_{xx}(\omega)]$ \cite{Hosur2012}, and a result with the appropriate logarithmic divergence for $\Im[\sigma_{xx}(\omega)]$ \cite{Rosenstein2013}.
This can be understood in terms of charge renormalization due to the band structure, but for ease of our purposes we let $\Lambda \sim 1/a_0$ where $a_0$ is the lattice spacing.
For our plots we choose a lattice spacing of $a_0=\unit[1]{nm}$, a thickness of $d=\unit[20]{nm}$, $b=0.3 (2\pi/a_0)$, $\Lambda=2\pi/a_0$, $v_{\mathrm{F}} = \unit[6\times 10^5]{m/s}$, and $\mu=0$ unless its the parameter we are varying.  

Now, one can use one of two equivalent ways of calculating the Casimir energy: the reflection matrix as given in Ref.~\cite{Tse2012}, or using a microscopic analysis to find the photon dressed RPA current-current correlators \cite{supplement}.
In order to avoid an unphysical negative $\sigma_{xx}(i\omega)$ as well as for consistency, we cutoff the photon energies in the Lifshitz formula to run from $0$ to $\Lambda$---an approximation valid for $a\gg \frac{c}{v_{\mathrm{F}}} \Lambda^{-1}$.

First, we see that we get an anti-trap for these at approximately $\unit[650]{nm}$, and if we increase $b$ as in Fig.~\ref{fig:WeylVaryb} (with, say, an applied magnetic field), it not only moves closer to zero separation, but the overall repulsive behavior can be enhanced.
On the other hand, if we increase $v_{\mathrm F}$ as we see in Fig.~\ref{fig:WeylVaryvf}, we see the region of attraction is suppressed, but the overall repulsive behavior at long distances is maintained.
Modifying $\Lambda$ will have effects similar to modifying $v_{\mathrm F}$, but since it appears logarithmically, it would need to change by orders of magnitude to give appreciable changes (a simple plot for this is provided in the Supplemental Material \cite{supplement} but is not relevant for the discussion here).
This antitrap effect is occurring at short distances when higher order band effects also play a role, but any other effects will contribute to the longitudinal conductivity in such a way that an anti-trap will appear.

Interestingly, when we introduce a finite chemical potential as we see in Fig.~\ref{fig:WeylVarymu}, in addition to the anti-trap we get at shorter distances, we start to see a trap at much longer distances appear.
This is not surprising since at zero frequency there is a divergent longitudinal conductivity.
Thus, we know that at long distances, the Casimir force must be attractive, but by modifying the Hall effect, we have an intermediate regime of repulsion.
A similar effect would occur if we took finite temperatures or disorder corrections to the longitudinal conductivities.

Considering the form of the conductance in terms of the fine-structure constant $\sigma_{xy}d/c = \alpha \frac{2bd}\pi$, we see that $b d$ controls the strength of the repulsion in the thin-film limit. 
Without longitudinal conductance, the repulsive regime roughly corresponds to when $(\sigma_{xy}d/c)^2 \lesssim \sigma_{xy}d/c$ or equivalently $ \frac{2bd}\pi\lesssim \frac1{\alpha}$. 
The longitudinal conductance introduces $v_\mathrm{F}$ into the scheme, relevant photons have $\omega \approx c/a$, and thus it becomes important for $\sigma_{xx} d/c \sim \alpha \frac{c}{v_\mathrm{F}} \frac{d}{a} \gtrsim O(1)$ (neglecting constants) which both emphasizes that $v_\mathrm{F}$ controls the longitudinal conductance's contribution to the Casimir effect and that the term is suppressed at longer distances.


We have shown here how Weyl semimetals can exhibit a tunable repulsive Casimir force (with, for instance, magnetic-field tuning $\mathbf b$) and how it can depend on the thickness of the material. In the thin-film limit, we showed how the semimetallic nature of these materials can work to create attraction at shorter distance scales and how a finite longitudinal conductivity will create long-distance attraction along with repulsion at intermediate distances.
Recently the first experimental observation of Weyl semimetals \cite{Huang2015,Zhang2015} provided optimism that these theoretical materials could be a reality.
The marginal nature of these materials could be useful for controlling the Casimir force between attractive and repulsive regimes.

This work was supported by the DOE-BES (Grant No. DESC0001911) (A.A.A. and V.G.), the JQI-PFC (J.H.W.), and the Simons Foundation. We thank Liang Wu and Mehdi Kargarian for discussions.

\bibliography{arxivreferences}

\onecolumngrid

\hspace{12pt}
\hrule

\section*{Supplementary materials}

 \section{Axionic Electrodynamics}

In the main text, we mention the axionic term which appears in the action alongside the usual Maxwell action
\begin{align}
\label{eq:axionic-actionsup}
  S_{A} = \frac{e^2}{32 \pi^2\hbar c} \int d^3 r\, dt \, \theta(\mathbf r, t)  \epsilon^{\mu \nu \alpha \beta} F_{\mu \nu} F_{\alpha \beta},
\end{align}
where $e$ is the electric charge, $\hbar$ is Planck's constant, $c$ is the speed of light, $F_{\mu\nu} = \partial_\mu A_\nu - \partial_\nu A_\mu$ is the electromagnetic field tensor, $\epsilon^{\mu \nu \alpha \beta}$ is the fully antisymmetric 4-tensor, and $\theta(\mathbf r, t) = 2 \mathbf b \cdot \mathbf r - 2 b_0 t$ is the axionic field.

For our purposes, we will set $\mathbf b = b \hat{\mathbf z}$ and $b_0 = 0$.
If we only apply $S_{\text{A}}$ for $z>0$, there is no resulting surface current (i.e. this is the surface without Fermi arcs), and the current response is
\begin{align}
  j^x(\mathbf r) & = \frac{e^2 b}{2 \pi^2} E^y(\mathbf r), \\
  j^y(\mathbf r) & = - \frac{e^2 b}{2 \pi^2} E^x(\mathbf r).
\end{align}

Now, we solve Maxwell's equations in the bulk Hall system after taking the Fourier transform
\begin{align}
  \mathbf k \cdot \mathbf E & = -i \sigma_{xy} \hat{\mathbf z} \cdot \mathbf B, \\
  \mathbf k \cdot \mathbf B & = \omega \mathbf B, \\
  \mathbf k \times \mathbf E & = \omega \mathbf B, \\
  \mathbf k \times \mathbf B & = i \sigma_{xy} \hat{\mathbf z} \times \mathbf E - \omega \mathbf E.
\end{align}	
One can define a frequency-dependent dielectric permitivity $\epsilon(\omega)$ to be
\begin{align}
  \epsilon(\omega) & = \begin{pmatrix}
    1 &  i \sigma_{xy}/\omega & 0 \\
    - i \sigma_{xy}/\omega & 1 & 0 \\
    0 & 0 & 1
  \end{pmatrix}
\end{align}
in which case, Maxwell's equations can be recast as a single equation for the electric displacement $\mathbf D = \epsilon(\omega) \mathbf E$,
\begin{align}
\label{eq:Maxwell-displacement-fieldsup}
  [\mathbf k \otimes \mathbf k - k^2 \mathbb I] \epsilon^{-1}(\omega) \mathbf D = - \omega^2 \mathbf D.
\end{align}
The determinant of this matrix equation yields the frequencies that a wave vector $\mathbf k$ can have.
In our case, we obtain
\begin{align}
\label{eq:Bulk-weyl-dispersionsup}
  \omega_{\pm}^2 = k^2 + \tfrac12 \sigma_{xy}^2 \pm \sigma_{xy} \sqrt{ k_z^2 + \tfrac14 \sigma_{xy}^2}.
\end{align}
There is a polarization associated with each of these frequencies which we can obtain from Eq.~\eqref{eq:Maxwell-displacement-fieldsup} using $\mathbf k \cdot \mathbf D = 0$ (no free charge).
We choose $\mathbf e_1 = \hat{\mathbf y} \times \hat{\mathbf k}$ and $\mathbf e_2= \hat{\mathbf y}$ as our basis for the polarizations (assuming $k_y = 0$ without loss of generality).
The resulting (unnormalized) polarizations are
\begin{align}
  \mathbf D_{1,2} = \frac{\omega_{\pm}}{k}\left(\sqrt{k_z^2 + \tfrac14 \sigma_{xy}^2} \mp \tfrac12 \sigma_{xy} \right) \mathbf e_1 \pm i k_z \mathbf e_2.
\end{align}
Notice that these polarizations are elliptical.

To find the reflection of a wave off a half space filled with this material, $k_x$, $k_y$, and $\omega$ must remain the same on either side of the material, but $k_z$ can change, and matching both sides of the dispersion Eq.~\eqref{eq:Bulk-weyl-dispersionsup}, we obtain simply that for an incident wave with wave-vector $\mathbf q = (q_x, q_y, q_z)$, the transmitted wave has
\begin{align}
  (k_z^{\pm})^2 = q_z(q_z \pm \sigma_{xy}). \label{eq:kz-transmitsup}
\end{align}
Each of these can be associated with a (unnormalized) polarization as (assuming $q_y = 0$ without loss of generality)
\begin{align}
  \mathbf D_{\pm}= \tfrac{\omega}{k^{\pm}} (q_z \pm \sigma_{xy}) \mathbf{e}_1 \mp i k_z^\pm \mathbf{e}_2.
\end{align}
At this point, we note some interesting electromagnetic properties appearing here.
Eq.~\eqref{eq:kz-transmitsup} implies that this material is birefringent, and for an incident wave (at any angle) with $q_z< |\sigma_{xy}|$, only one (elliptical) polarization even propagates into the material while the other is an evanescent wave -- independent of the angle of incidence.

\subsection{Reflection coefficient of semi-infinite bulk}

Now, to obtain the reflection matrix, we call our incident wave $\mathbf E_0$ with wave-vector $\mathbf q$, our reflected wave $\mathbf E_r$ with wave-vector $\mathbf q_r =(q_x,q_y,-q_z)$, as well as $\mathbf E_{\pm} = \epsilon^{-1}(\omega) \mathbf D_\pm$ with wave vectors $\mathbf k_\pm$ for the two polarizations it is transmitted into.
The relevant Maxwell equations at the interface between vacuum and the bulk Hall material are then given by
\begin{align}
\begin{split}
  (\mathbf E_0 + \mathbf E_r - \mathbf E_+ - \mathbf E_-) \times \hat{\mathbf z} & = 0, \\
  (\mathbf q \times \mathbf E_0 + \mathbf q_r \times \mathbf E_r - \mathbf k_+ \times \mathbf E_+ - \mathbf k_- \times \mathbf E_-) \times \hat{\mathbf z} & = 0.
\end{split} \label{eq:matching-conditionssup}
\end{align}
We can break up the polarization of the incident and reflected waves into transverse electric (TE) and transverse magnetic (TM), and the reflection matrix is defined such that
\begin{align}
  \begin{pmatrix}
  E^{\mathrm{TM}}_r \\ E^{\mathrm{TE}}_r
  \end{pmatrix} & = R(\omega, \mathbf q)\begin{pmatrix}
  E^{\mathrm{TM}}_0 \\ E^{\mathrm{TE}}_0
  \end{pmatrix}.
\end{align}
Solving for this matrix, we obtain
\begin{align}
R_\infty(\omega, \mathbf q) = \frac1{ \sigma_{xy}} \begin{pmatrix}
 \sigma_{xy} + k_z^- - k_z^+ & i (2 q_z - k_z^- - k_z^+) \\
 -i (2 q_z - k_z^- - k_z^+) & \sigma_{xy} + k_z^- - k_z^+ 
\end{pmatrix}.
\end{align}
This can then be rotated to imaginary frequencies and the result is in the main text.

\subsection{Reflection coefficient of thickness $d$ sample}

If we have a material of thickness $d$, then we need additional matching conditions due to Maxwell's equations at the other interface.
This requires restricting our action Eq.~\eqref{eq:axionic-action} to $0<z<d$.
To solve this, we just need to add another set of matching conditions.
In addition to the incident $\mathbf E_0$ and reflected $\mathbf E_r$ waves, we now have forward moving Weyl polarizations $\mathbf E_\pm^{\uparrow}$ with $\mathbf k_\pm^\uparrow = (k_x,k_y,k_z^\pm)$, backwards moving Weyl polarizations $\mathbf E_\pm^{\downarrow}$ with $\mathbf k_\pm^\downarrow = (k_x,k_y, -k_z^\pm)$, and a transmitted wave $\mathbf E_t$ with wave-vector the same as the transmitted $\mathbf k$.

The resulting matching conditions are
\begin{align}
\begin{split}
  (\mathbf E_0 + \mathbf E_r - \mathbf E^\uparrow_+ - \mathbf E^\uparrow_- - \mathbf E^\downarrow_+ - \mathbf E^\downarrow_- )\times \hat{\mathbf z} & = 0, \\
  (\mathbf k \times \mathbf E_0 + \mathbf k_r \times \mathbf E_r - \mathbf k^\uparrow_+ \times \mathbf E^\uparrow_+ - \mathbf k^\uparrow_- \times \mathbf E^\uparrow_- - \mathbf k^\downarrow_+ \times \mathbf E^\downarrow_+ - \mathbf k^\downarrow_- \times \mathbf E^\downarrow_-)\times \hat{\mathbf z} & = 0, \\
  (\mathbf E^\uparrow_+ e^{i k_z^+ d} + \mathbf E^\uparrow_- e^{i k_z^- d}+ \mathbf E^\downarrow_+ e^{-i k_z^+ d} + \mathbf E^\downarrow_- e^{-i k_z^- d} - \mathbf E_t e^{i k_z d})\times \hat{\mathbf z} & = 0,\\
  (\mathbf k_+^\uparrow \times \mathbf E^\uparrow_+ e^{i k_z^+ d} + \mathbf k_-^\uparrow \times \mathbf E^\uparrow_- e^{i k_z^- d} \phantom{+ \mathbf k_-^\downarrow \times \mathbf E^\downarrow_- e^{-i k_z^- d} - \mathbf k \times \mathbf E_t e^{i k_z d})\times \hat{\mathbf z}} \\+ \mathbf k_+^\downarrow \times \mathbf E^\downarrow_+ e^{-i k_z^+ d} + \mathbf k_-^\downarrow \times \mathbf E^\downarrow_- e^{-i k_z^- d} - \mathbf k \times \mathbf E_t e^{i k_z d})\times \hat{\mathbf z} & = 0.
\end{split}
\end{align}
These equations can still be solved and the result is
\begin{align}
  R_d(\omega, \mathbf q) = \begin{pmatrix}
    R_{xx} & R_{xy} \\ -R_{xy} & R_{xx}
  \end{pmatrix},
\end{align}
where
\begin{align}
  R_{xx} & = \sigma_{xy}  \{\sin (k_z^- d) [i k_z^+ \cos (k_z^+ d)+\sigma_{xy}  \sin (d
   k_z^+)]-i k_z^- \cos (k_z^- d) \sin (k_z^+ d)\}/D, \\
   R_{xy} & = \sigma_{xy} 
   \{k_z^- \cos (k_z^- d) \sin (k_z^+ d)+\sin (k_z^- d) [k_z^+
   \cos (k_z^+ d)-2 i k_z \sin (k_z^+ d)]\}/D, \\
   D = & [2 i k_z^-
   \cos (k_z^- d)+(2 k_z-\sigma_{xy} ) \sin (k_z^- d)] [2 i k_z^+ \cos
   (k_z^+ d)+(2 k_z+\sigma_{xy} ) \sin (k_z^+ d)].
\end{align}
And again, this can be rotated to imaginary frequencies to obtain the result in the main text.

And as stated in the text, the various limits (semi-infinite to thin film limits) apply to this reflection matrix.

\section{Calculating the conductivities}

In order to calculate the conductivities in the clean limit we consider the Hamiltonian near a Weyl node
\begin{align}
  H_W = \pm \hbar v_{\mathrm{F}} \bm \sigma \cdot (\mathbf k \pm \mathbf b).
\end{align}
For simplicity we set $\hbar = 1 = v_{\mathrm{F}}$ unless otherwise specified.
We consider a pair of these nodes, calculating the quantities separately for each node and adding them together (which will just introduce a factor of two for both $\sigma_{xx}$ and $\sigma_{xy}$).

To find the conductivities, we find a complete basis of states which are easily found by diagonalizing the two-by-two matrix $H_W$ in momentum space (label them $\ket{f_{k\pm}}$).
The relevant matrix elements are then (choosing the negative sign for the Hamiltonian)
\begin{align}
  \braket{f_{k+}|\sigma_x|f_{k-}} & = \frac{(k_z - b)k_x + i \epsilon k_y}{\epsilon \sqrt{\epsilon^2 - (k_z-b)^2} }, \\
  \braket{f_{k+}|\sigma_y|f_{k-}} & = \frac{(k_z - b)k_y - i \epsilon k_x}{\epsilon \sqrt{\epsilon^2 - (k_z-b)^2} }, \\
  \braket{f_{k\pm}|\sigma_x|f_{k\pm}} & = \pm \frac{k_x}{\epsilon},
\end{align}
where $\epsilon= \sqrt{k_x^2 + k_y^2 + (k_z-b)^2} $.

We then fix $k_z$ and use the Kubo-Greenwood formula for the intra- and inter-band transitions separately to obtain two-dimensional conductivities. Thus,
\begin{align}
  \tilde\sigma_{\mu \nu}^{\mathrm{inter}}(i\omega;k_z)= \frac{e^2}{i}\sum_{\mathbf k, \gamma\neq \gamma'} \frac{n_{\mathbf k \gamma} - n_{\mathbf k \gamma'}}{\epsilon_{\mathbf k \gamma }- \epsilon_{\mathbf k \gamma'}}  \frac{\braket{f_{\mathbf k \gamma} | j_{ \mu} | f_{\mathbf k \gamma'}}\braket{f_{\mathbf k \gamma'} | j_{ \nu} | f_{\mathbf k \gamma}}}{i \omega + \epsilon_{\mathbf k \gamma} - \epsilon_{\mathbf k \gamma'}},
\end{align}
where $n_{\mathbf k \gamma}$ is the occupation in that band of that momentum and $j_\mu = \sigma_\mu$ are the single particle current operators.
For intra-band quantities with $n_{\mathbf k \pm} = \theta(\mu \mp \epsilon_{\mathbf k})$
\begin{align}
  \tilde\sigma_{\mu \nu}^{\mathrm{intra}}(i\omega;k_z)= - \frac{e^2}{i}\sum_{\mathbf k } \delta(\mu - \epsilon_{\mathbf k,+})  \frac{\braket{f_{\mathbf k +} | j_{ \mu} | f_{\mathbf k +}}\braket{f_{\mathbf k +} | j_{ \nu} | f_{\mathbf k +}}}{i \omega },
\end{align}
assuming only the upper-band for simplicity and without loss of generality (due to particle-hole symmetry).

Adding these contributions together at finite chemical potential yields \cite{Tse2012} 
\begin{align}
  \tilde\sigma_{xx}( i \omega; k_z) &= \frac{e^2}{4\pi} \left[\left(1 - \frac{4 (k_z - b)^2}{\omega^2} \right) \frac{i}4 \log\left( \frac{2\Delta - i \omega}{2\Delta + i\omega}\right) + \frac{\Delta}{  \omega}\right], \\
  \tilde\sigma_{xy}( i\omega; k_z ) & = \frac{e^2}{2\pi} \left[ \frac{k_z - b}{\omega} \frac{i}2 \log\left( \frac{2\Delta - i \omega}{2\Delta + i\omega}\right)\right],
\end{align}
where $\Delta = \max\{|k_z - b|,|\mu|\}$.
With these quantities we can then use the cutoff procedure explained in \cite{Goswami2013}
\begin{align}
  \sigma_{\mu \nu}(i\omega) = \int_{-\Lambda}^\Lambda \frac{d k_z}{2\pi} \tilde \sigma_{\mu \nu}(i\omega; k_z).
\end{align}
Through which we obtain (throwing away terms that go to zero as the cutoff increases to infinity and multiplying by two for the two nodes and bringing back in the constants $\hbar$ and $v_{\mathrm{F}}$)
\begin{align}
 \sigma_{xx}(i\omega) & = \frac{e^2}{12 \pi^2 \hbar v_{\mathrm F}} \left[ \tfrac 53 \omega + 2 \omega \log \left( \tfrac{v_{\mathrm F} \Lambda}{\omega} \right) + 4 \tfrac{\mu^2}{\hbar^2\omega} - \omega \log \left(1 + \tfrac{4 \mu^2}{\hbar^2\omega^2} \right)\right], \\
 \sigma_{xy}(i \omega) & = \frac{e^2 b}{2 \pi^2 \hbar}.
\end{align}
These quantites are what we use in the next section as input for the Casimir Force. \label{sec:conductivitysup}

\section{Casimir Force calculation}

\subsection{Differentiating the Casimir energy}

In the main text, we find the Casimir pressure by taking the derivative of the energy $P_c = - \partial E_c/\partial a$. This leads to the expression for force found in the text for two semi-infinite Weyl plates
\begin{align}
P_c = \frac{2 \hbar c}{(2\pi)^2} \int d q_z\, q_z^3 \; g\!\left[\tfrac{q_z}{\sigma_{xy}/c}, 2 q_z a \right], \label{eq:semi-inf-forcesup}
\end{align}
and the function $g(u,v)$ is defined by
\begin{align}
  g(u,v) = -4 \frac{ R_{xx}(u)^2 - R_{xy}^2(u) - [R_{xx}(u)^2 + R_{xy}^2(u)]^2 e^{-v}}{2[R_{xx}(u)^2 - R_{xy}^2(u)] - [R_{xx}(u)^2 + R_{xy}^2(u)]^2 e^{-v}},
\end{align}
where $R_{xx}(u)$ and $R_{xy}(u)$ are the matrix elements of $R_\infty(ic q_z/\sigma_{xy})$ (Eq.~(3) in the main text),
\begin{align}
 R_{xx}(u) & =\sqrt{2u(\sqrt{u^2+1}-1)} - 1, \\
 R_{xy}(u) & = 2u - \sqrt{2u(\sqrt{u^2+1}+1)}.
\end{align}

\subsection{Two-dimensional plates calculation}

Another well-known approach completely equivalent to the Lifshitz formula comes directly from quantum field theory.

To put it briefly, if we have a conductivity like we calculated above and write it as a response function $\Pi(i\omega)$ such that
\begin{align}
\sigma(i\omega) &= -\Pi(i\omega)/\omega
\end{align}
And calculate the RPA response function considering photons ``skimming'' along the surface of our material
\begin{align}
\widetilde{\Pi}(i\omega,q) &= \left[\mathbb I - \Pi(i\omega) D(i\omega,q,0) \right]^{-1} \Pi(i\omega),
\end{align}
with the photon propagator
\begin{align}
D(i\omega,q,z) &= \begin{pmatrix} \frac{\sqrt{q^2+\omega^2}}{2\omega^2} & 0 \\ 0 & \frac{1}{2\sqrt{q^2+\omega^2}}\end{pmatrix} e^{-\sqrt{q^2+\omega^2}|z|},
\end{align}
then the Casimir energy takes the form
\begin{align}
E_c(a) &= \frac1\pi \int \frac{d^2q}{(2\pi)^2} \int _0^{\infty} d\omega \tr \log\left[\mathbb I - \widetilde{\Pi}_1(i\omega,q)D(i\omega,q,a)\widetilde{\Pi}_2(i\omega,q)D(i\omega,q,a) \right],
\end{align}
For our particular case of the conductivities introduced in Section \ref{sec:conductivitysup}, we need to also cutoff the frequencies in this integral to go from $0$ to $v_{\mathrm F}\Lambda$ for consistency.
We expect higher energy virtual photons to not play a large role.

\begin{figure}
  \includegraphics[width=8cm]{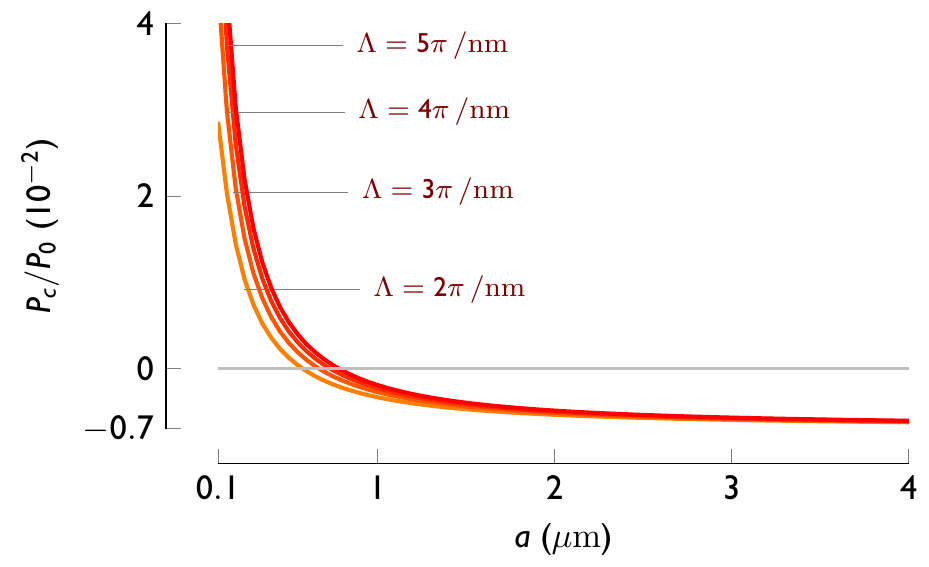}
  \caption{The Casimir energy for the conductivities as defined in the main text with $b= \unit[0.3 (2\pi)]{nm^{-1}}$, $v_{\mathrm{F}} = \unit[6\times 10^5]{m/s}$, and $\mu=0.$ The cutoff is varied here.}
  \label{fig:cutoff-varysup}
\end{figure}

We also show the result we obtain from varying the cutoff in Fig.~\ref{fig:cutoff-varysup}. 
Note that it does not affect the effect much unless it varies by orders of magnitude.

\end{document}